\documentclass[aps,pra,showpacs,twocolumn,superscriptaddress]{revtex4}

\usepackage{epsfig}
\usepackage{amsmath,amssymb,amsthm}
\usepackage{graphicx}
\usepackage{psfrag}
\usepackage{bm}

\def\s2{\sigma^2}

\begin{document}
\title{Fractional revivals through R\'enyi uncertainty relations}

\author{E. Romera}
\affiliation{
Instituto Carlos I de F{\'\i}sica Te\'orica y
Computacional, Universidad de Granada, Fuentenueva s/n, 18071 Granada,
Spain}
\affiliation{Departamento de F\'isica At\'omica, Molecular y Nuclear, 
Universidad de Granada, Fuentenueva s/n, 18071 Granada, Spain
}

\author{F. de los Santos}
\affiliation{
Instituto Carlos I de F{\'\i}sica Te\'orica y
Computacional, Universidad de Granada, Fuentenueva s/n, 18071 Granada,
Spain}
\affiliation{Departamento de Electromagnetismo y F{\'\i}sica de la
Materia, Universidad de Granada, Fuentenueva s/n, 18071 Granada,
Spain}

\date{\today}

\begin{abstract}
We show that the R\'enyi uncertainty relations give a good description of the dynamical behavior of 
wave packets and constitute a sound approach to revival phenomena by analyzing
three model systems: the simple harmonic oscillator, the infinite square well, and
the quantum bouncer.
We prove the usefulness of entropic uncertainty relations as a tool for identifying fractional
revivals by providing a comparison in different contexts with the with the usual Heisenberg uncertainty relation 
and with the common approach in terms of the autocorrelation function. 

\end{abstract}
\pacs{03.65.Ge;03.65.Sq}
\maketitle

\section{Introduction}

The temporal evolution of wave packets displays a wide variety of 
non-classical effects and, in this regard, revivals and fractional revivals 
have raised great interest during the last years \cite{1}.
Revivals occur when a wave packet solution of the Schr\"odinger 
equation returns to a state that closely reproduces its initial wave form,
at multiples of a revival time $T_{\rm rev}$. 
Fractional revivals appear as the temporal self-splitting of the 
initial wave packet into a collection of a number of scaled copies.
Assuming that the initial state is a superposition of eigenstates $u_n(x)$ 
sharply peaked around some $n_0$, revival times can then be obtained from the 
Taylor series of the energy spectrum $E_n$ around $E_{n_0}$
\begin{equation}
E_n \approx E_{n_0}+E'_{n_0} (n-n_0)+\frac{E''_{n_0}}{2} (n-n_0)^2 +\ldots
\end{equation}
as $T_{\rm rev}=2h/|E''_{n_0}|$. Fractional revival times 
are given by rational fractions of the revival time \cite{1b,1c,rob}.
These phenomena have been observed in many experimental situations and 
studied theoretically in a variety of quantum
systems such as Rydberg wave packets in atoms and molecules, Bose-Einstein
condensates, etc \cite{1,1b,1c,rob,yeazell,exp}. Interestingly, the phenomenon 
of revivals is at the basis of a method for isotope separation \cite{isotope}
as well as for number factorization \cite{factorization}.

Revivals and fractional revivals are usually quantified using the autocorrelation function
\begin{equation}
A(t)\equiv\int_{-\infty}^{\infty}
\psi^{*}(x,t)\psi(x,0) dx =\int_{-\infty}^{\infty} \phi^{*}(p,t)\phi(p,0) dp,
\label{autocorrelation}
\end{equation}
which is the overlap between the initial and the time-evolving wave
packet in either the position or momentum representation. 
Given an initial state, $A(t)$ decreases in time and 
the occurrence of revivals reflects in the return of $A(t)$ to its initial value of unity, 
or in the appearance of relative maxima in the case of fractional revivals. 
Revival phenomena have been studied as well by tracking 
the time evolution of the expectation values 
of some quantities \cite{rob,sun,don}, and an approach
based on a finite difference eigenvalue method has been recently
put forward from which the various orders of revivals can be 
directly calculated rather than searching for them \cite{laserphysics}.

Recently, the sum of Shannon information entropies in position 
and momentum conjugate spaces has been shown to provide a
useful tool for describing fractional revivals, complementary 
to the usual approach in terms of the autocorrelation function \cite{nuestro}. 
The underlying idea is that the position-space Shannon entropy measures the uncertainty in
the localization of the particle in space \cite{uffink}, so the lower
is this entropy the more concentrated is the probability density $|\psi(x,t)|^2$, the
smaller is the uncertainty, and the higher is the accuracy in
predicting the localization of the particle. Equivalently, the momentum-space
entropy measures the uncertainty in predicting the momentum of the
particle. 
Thus, the sum of Shannon entropies gives an account of the spreading 
(high entropy values) and the regenerating (low entropy values) of initially well localized 
wave packets during the time evolution,
and the temporary formation of fractional revivals of the wave function is given by
the relative minima of the sum of Shannon information entropies in both conjugate spaces
due to the fact that the uncertainty relation is
saturated only for Gaussian wave packets \cite{nuestro}.

In this article, we expand on the entropic uncertainty approach by describing fractional 
revivals by means of the sum of the R\'enyi entropies in position and momentum conjugate spaces,
and make a comparison with an analysis based on the time evolution of the standard 
uncertainty relation in terms of the variance of the probability density in both position and 
momentum spaces. In the next section we review some basic properties
of the R\'enyi entropy and apply it to three specific model systems: the harmonic oscillator, which 
can be worked out analytically; the infinite square well, which exhibits perfect revival behavior,
and the quantum bouncer, which has been recently realized using neutrons \cite{bouncer1} and
atomic clouds \cite{bouncer2}

\section{Fractional revivals and the R\'enyi uncertainty relation}

The R\'enyi entropy is a generalization of the Shannon entropy and has been widely 
employed in the study of quantum systems. Among others applications, 
it was used in the analysis of quantum entanglement \cite{cite14}, 
quantum communication protocols \cite{7}, localization properties of
Rydberg states and spin systems \cite{cite1416}, and quantum measurement and
decoherence \cite{cite89}. It is defined in terms of a general probability
density $f(x)$ as (see \cite{biarenyi} and references therein)
\begin{equation}
R_{f}^{(\alpha)}\equiv
\frac{1}{1-\alpha}\ln \int_{-\infty}^{\infty}\left[f(x)\right]^{\alpha}dx \quad
\text{for  } 0<\alpha<\infty\quad \alpha\neq 1.
\label{definition}
\end{equation} 
In terms of the probability density in position and momentum spaces, $\rho(x)=|\phi(x)|^2$
and $\gamma(p)=|\phi(p)|^2$, respectively, the R\'enyi uncertainty relation is given by
\cite{biarenyi}
\begin{equation}
 R_{\rho}^{(\alpha)} +
 R_{\gamma}^{(\beta)}\geq -\frac{1}{2(1-\alpha)}\ln\frac{\alpha}{\pi}-\frac{1}{2(1-\beta)}\ln\frac{\beta}{\pi}
\label{uncer}
\end{equation} 
where $1/\alpha+1/\beta=2$. 
From the continuous or integral R\'enyi entropy, as defined in equation (\ref{definition}),
it is straightforward to obtain the following limiting behaviors
\begin{eqnarray}
R_f^{(\alpha)} &\xrightarrow[\alpha \to {}1]\,& S_f\equiv -\int f(x)\ln f(x)dx, \nonumber \\ 
R_f^{(\alpha)} &\xrightarrow[\alpha \to {}\infty]\,&-\ln [\max_x f(x)].
\end{eqnarray}
Consequently, in the limits $\alpha\rightarrow 1$ and $\beta \rightarrow 1$ 
the R\'enyi uncertainty relation (\ref{uncer}) reduces to Shannon's \cite{BBM},
\begin{equation}
S_\rho+S_\gamma \geq 1+\ln (\pi)
\label{B}
\end{equation}
which can thus be considered a particular case of the former. 
Clearly, all of the above discussion on the benefits of using the Shannon entropy 
to study revival phenomena carries over to the R\'enyi entropy as well. 
It is noteworthy, however, that at the core of the success of the entropic approach is the 
existence of the uncertainty relations (\ref{uncer}) and (\ref{B}).
For this same reason, it is expected that insights on fractional revivals can also be gleaned
from the standard variance-based Heisenberg uncertainty relation,
which in the case of the usual position and momentum variables leads to the celebrated inequality
\cite{kennard}
\begin{equation}
\Delta x \Delta p\geq \frac{\hbar}{2},
\label{usual_uncer}
\end{equation}
with $(\Delta x)^2=\langle x^2 \rangle-\langle x\rangle^2$ 
and $(\Delta p)^2= \langle p^2 \rangle-\langle p \rangle^2$.
Note that 
inequalities (\ref{uncer}) and (\ref{B}) are saturated by Gaussian probability
distributions. 

Next, we make use of the relations (\ref{uncer}) and (\ref{usual_uncer}) to 
study revival and fractional revivals in three illustrative cases.

\subsection{Simple harmonic oscillator}

The harmonic oscillator provides the most straightforward example of a system showing bound
states for which the periodic motion of wave packet solutions (especially
Gaussian) is easily derivable. 
In the present case $V(x)=\frac{1}{2}m \omega^2 x^2$ and all wave packet solutions, Gaussian
or not, satisfy $\psi(x,t+mT_{cl})=(-1)^m \psi(x,t)$, where $T_{cl}=2\pi/\omega$ is the classical period.
The associated time dependent expectation values $\langle x \rangle_t$ and $\langle p \rangle_t $ 
behave exactly as for a classical particle \cite{rob,styer},
\begin{eqnarray}
\langle x \rangle_t &=&  \langle x \rangle_0 \cos( \omega t)+\frac{\langle p \rangle_0 }{m\omega} \sin(\omega t), \nonumber \\
\langle p \rangle_t &=& -m \omega^2 \langle x \rangle_0 \sin(\omega t) + \langle p \rangle_0 \cos(\omega t).
\end{eqnarray}
The behavior of the uncertainties can also be worked out explicitly. For our purposes, it suffices
to focus on initial Gaussian wave functions, also known as squeezed states,
\begin{equation}
\psi(x,0)=\frac{1}{\sqrt{\sigma\sqrt{\pi}}} e^{i p_0 x/\hbar} e^{-(x-x_0)^2/2\sigma^2}.
\label{initial}
\end{equation}
The time-evolved wave function is then given in closed form for arbitrary times as \cite{rob} 
\begin{equation}
\psi(x,t)=\frac{1}{\sqrt{|L(t)|\sqrt{\pi}}} e^{\frac{S(x,t)}{2\sigma L(t)}}
\end{equation}
with 
\begin{equation}
L(t)=\sigma \cos(\omega t)+\frac{i\hbar}{m\omega \sigma}\sin(\omega t)
\end{equation} 
and
\begin{eqnarray}
S(x,t)&=&
-x_0^2 \cos(\omega t) -\frac{2 x_0 p_0 \sin(\omega t)}{m\omega}-\frac{i \sigma^2 p_0^2 \sin(\omega t)}{m\omega \hbar} \nonumber \\
&&+2 \left(x_0 + \frac{i \sigma^2 p_0 }{\hbar}\right)x \nonumber \\
&&- \left[\cos(\omega t) + \frac{im\omega\sigma^2 \sin(\omega t)}{\hbar}\right] x^2,
\end{eqnarray}
from which the following position and momentum uncertainties can be computed
\begin{eqnarray}
\Delta x(t)  &=& |L(t)|/\sqrt{2}, \nonumber \\
\Delta p(t)  &=& \sqrt{\frac{(\hbar/\sigma)^2 \cos^2(\omega t)+ (m \omega \sigma)^2 \sin^2(\omega t)}{2}}. 
\end{eqnarray}
Thus, unless $\sigma^2=\hbar/m \omega$, both of them oscillate with a period equal to half of the natural
period of the classical simple harmonic oscillator \cite{rob,styer},
and so does their product $\Delta x \Delta p$. However, for the particular value 
$\sigma_{\rm coh}^2=\hbar/m \omega$ one finds a so-called coherent state 
for which the shape of the wave packet does not change in time. Consequently, its width also 
remains constant in time. One finds $\Delta x=\sigma_{\rm coh}/\sqrt{2}$ and $\Delta p= \hbar/(\sqrt{2}\sigma_{\rm coh})$
and therefore the product of the uncertainties is time independent and equals $\hbar/2$, what renders the coherent state a state
of minimum uncertainty. This phenomenology is of importance to several fields in quantum mechanics and
reflects in the behavior of the R\'enyi entropies in a clear way.
For the harmonic oscillator, the position and momentum R\'enyi entropies for parameters $\alpha$ and
$\beta$ can be readily calculated as,
\begin{eqnarray}
R_\rho^{(\alpha)}&=&\ln \left( \sqrt{\pi}|L(t)| \right)-\frac{\ln(\sqrt{\alpha})}{1-\alpha}, \\
R_\gamma^{(\beta)}&=&\ln \left(\frac{\sqrt{\pi}}{|L(t)|}\right)-\frac{\ln(\sqrt{\beta})}{1-\beta},
\end{eqnarray} 
and hence, when taken separately, they reproduce the classical periodic behavior
except for the particular value $\sigma_{\rm coh}^2$ for which the R\'enyi entropies are constants of the motion. 
In this case, moreover, their sum $R_\rho^{(\alpha)}+R_\gamma^{(\beta)}$ reaches the lower bound at all times, as  
expected for a system whose states remain Gaussian-like.

\begin{figure}
\includegraphics[width=8.5cm]{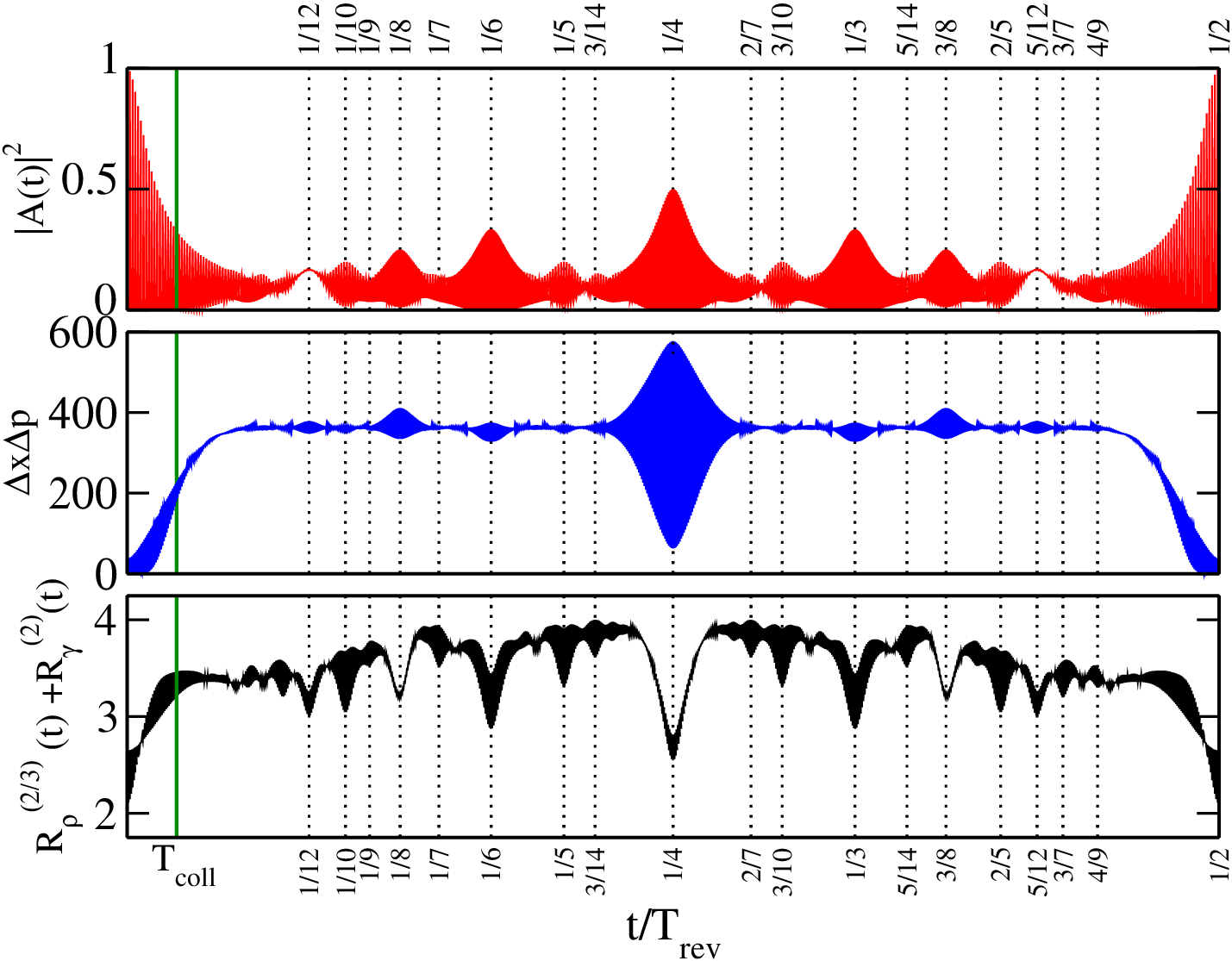}
\caption{(color online). 
Time dependence of (top panel, red curve) $|A(t)|^2$, (middle panel, blue curve)
$\Delta x(t) \Delta p(t)$,  and (bottom panel, black curve) $R_{\rho}^{(2/3)}(t)+R_{\gamma}^{(2)}(t)$ 
for an initial Gaussian wave packet with $x_0=L/2$, $p_0=400\pi$, and
$\sigma=\sqrt{2}/10$ in an infinite square-well.
The main fractional revivals are indicated by vertical dotted lines, and
the vertical, green solid line stands for the collapse time.
}
\label{fig1}
\end{figure}


\subsection{Infinite square well}

Consider a particle of mass $m$ in an infinite potential-well defined as
$V(x)=0$ for $0<x<L$ and $V(x)=+\infty$ otherwise.
The time-dependent wave function for a localized quantum wave packet is expanded as
a one-dimensional superposition of energy eigenstates as 
\begin{equation}
\psi(x,t)=\sum_n a_n u_n(x) e^{-i E_n t/\hbar},
\label{evolutionx}
\end{equation}
where the $u_n(x)$ represent the normalized eigenstates and $E_n$ the corresponding eigenvalues,
\begin{equation}
u_n(x)\sqrt{\frac{2}{L}} \sin \left(\frac{n\pi x}{L}\right), \qquad
E_n=\frac{n^2 \hbar^2\pi^2}{2 m L^2}.
\end{equation}
The infinite square well has exact revivals because the energy levels are integral multiples of a
common value (but not equally spaced). The classical and
revival periods are $T_{\rm cl}=2 m L^2/\hbar \pi n_0$ and $T_{\rm rev}=4mL^2/\hbar \pi$, 
respectively \cite{4,rob}. It is easy to see by direct substitution in equation (\ref{evolutionx})
that $\psi(L-x,t)=-\psi(x,0)$, so at a time $t=T_{\rm rev}/2$ the initial state reforms exactly, reflected around
the center of the well. This is the reason why the time span of the following analysis is $T_{\rm rev}/2$
rather than $T_{\rm rev}$.

We shall consider an initial Gaussian wave packet with a width $\sigma$, 
centered at a position $x_0$ and with a momentum $p_0$ as in equation (\ref{initial}). 
Assuming that the integration region can be extended to the whole real axis,
the expansion coefficients can be approximated with high accuracy by the
analytic expression

\begin{eqnarray}
a_n \approx \sqrt{\frac{4\sigma\pi}{L\sqrt{\pi}}} \frac{e^{-ip_0x_0/\hbar}}{2i} 
\Big[ e^{in\pi x_0/L} e^{-\sigma^2(p_0+n\pi \hbar/L)^2/2\hbar^2} \nonumber  \\ 
-e^{-in\pi x_0/L} e^{-\sigma^2(p_0-n\pi \hbar/L)^2/2\hbar^2} \Big].
\end{eqnarray}
To calculate the corresponding time dependent, momentum wave function
we use the Fourier transform of the equation (\ref{evolutionx}), that is,
\begin{equation}
\Psi(p,t)=\sum_n a_n \phi_n(p) e^{-i E_n t/\hbar}
\label{evolutionp}
\end{equation}
where
\begin{equation}
\phi_n(p)=\sqrt{\frac{\hbar}{\pi L}} \frac{p_n}{p^2-p_n^2}
\left[(-1)^n e^{ipL/\hbar}-1\right],
\end{equation}
with $p_n=\hbar \pi n /L$. 
Without loss of generality, we shall henceforth take $2m=\hbar=L=1$, 
$\sigma=\sqrt{2}/ 20$, and $x_0=L/2=0.5$ for the initial wave packet.
In our exemplary cases, we use $p_0=400\pi$ to obtain an appropriate
relative time scale. 

\begin{figure}
\includegraphics[width=8.5cm,angle=0]{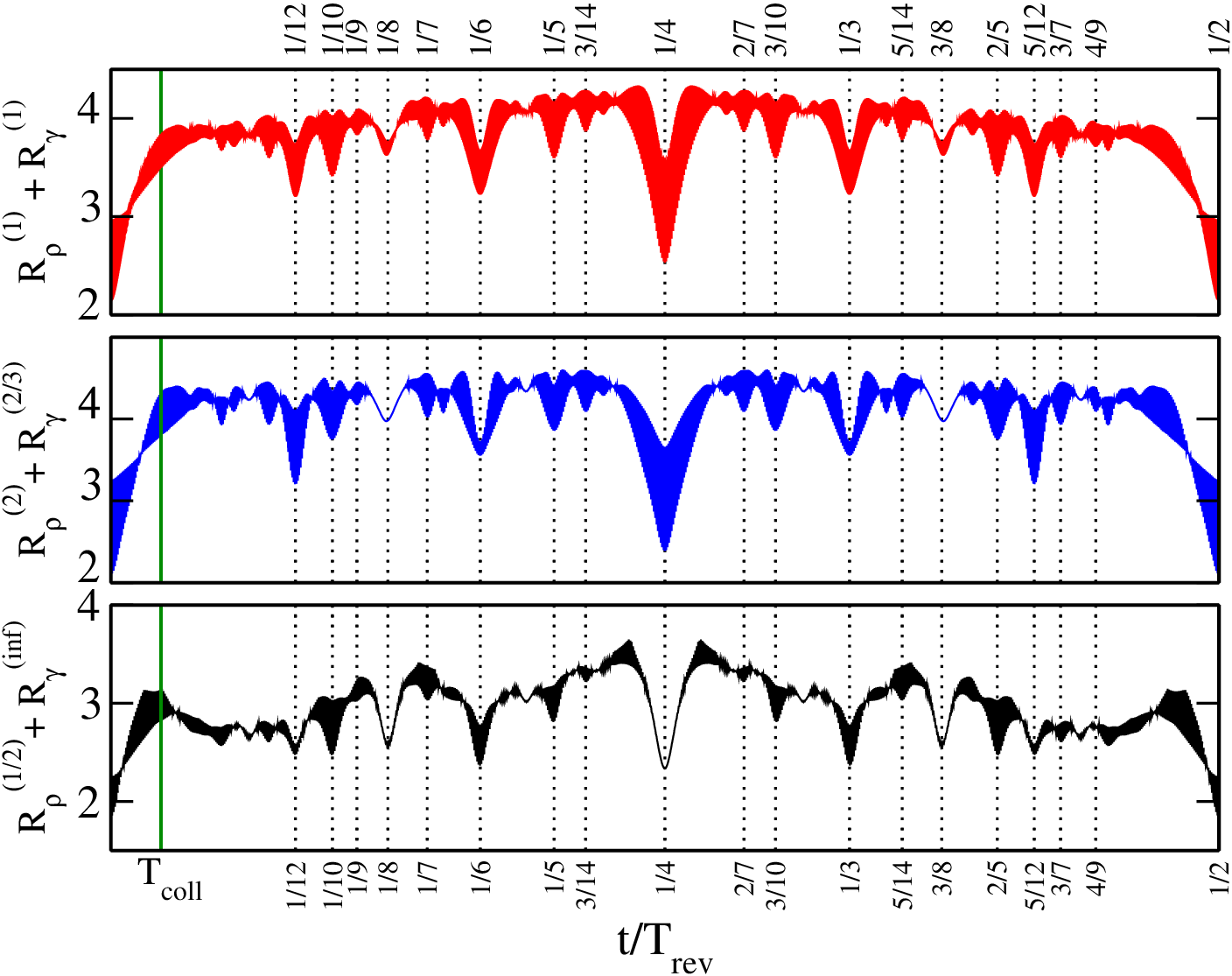}
\caption{(color online). 
Time dependence of (top panel, red curve) 
$R_{\rho}^{(1)}(t)+R_{\gamma}^{(1)}(t)$ (Shannon entropy),
(middle panel, blue curve) $R_{\rho}^{(2)}(t)+R_{\gamma}^{(2/3)}(t)$,
and (bottom panel, black curve) $R_{\rho}^{1/2}(t)+R_{\gamma}^{(\rm inf)}(t)$ 
for an initial Gaussian wave packet in an infinite square-well.  
The main fractional revivals are indicated by vertical dotted lines
and the vertical green curve stands for the collapse time.
Parameters as in Fig. \ref{fig1}.}
\label{fig2}
\end{figure}

In figure \ref{fig1} we show the manifestation of fractional
revivals with different tools: the usual autocorrelation function (top panel),
the expectation-value uncertainty product $\Delta x \Delta p$ (middle panel), 
and the sum of R\'enyi entropies with $\alpha=\frac{2}{3}$, $\beta=2$ (bottom panel).
At early times, the Gaussian wave packet evolves quasiclassically but in a few periods 
begins to delocalize and spreads almost uniformly across the entire well. This is the 
so-called collapse phase. The time-scale for this collapse has been estimated by means 
of an expectation value analysis to be \cite{rob,rob2}
\begin{equation}
T_{\rm coll}=\frac{1}{\sqrt{6}} \frac{mL \sigma}{\hbar}\simeq 0.0144
\end{equation}
for the set of parameters defined above. It is marked by the solid green line of figure \ref{fig1},
and coincides with the first maximum of the sum of R\'enyi entropies. By contrast, 
neither the autocorrelation function nor the curve for the position-momentum uncertainty
show any significant feature at $T_{\rm coll}$. 

At later times, the system undergoes a sequence of fractional revivals, the most important of 
which are shown by vertical dotted lines.
It can be observed that the sum of R\'enyi entropies accounts for them in the form of relative minima, 
only reaching the lower bound at the revival times. As shown in \cite{nuestro},
cases can be found within the infinite square well where the autocorrelation function 
fails to detect fractional revivals whereas the sum of entropies does, the
reason being that information entropies take into account the individual subpackets
the initial wave packet breaks up into, irrespective of their relative positions.
Additionally, it is shown here that the uncertainty relation for position and momentum also captures
the main fractional revivals, even though in a much less clear fashion (see, for instance, the 
fractional revivals occurring at $t=1/10$ or $t=1/5$). 
The time evolution of the uncertainty product was first used to study 
the evolution of wave packets in the context of revival phenomena in 
\cite{yeazell}. The fact that the expectation value analysis gives a poorer 
description than the entropic one suggests 
that the formulation of the uncertainty principle in term of R\'enyi entropies
is indeed stronger than that of the standard Heisenberg relation \cite{uffink,biarenyi}.

In contrast to the Shannon entropy, the R\'enyi entropy has a free parameter
and in figure \ref{fig2} it is shown how the fractional revivals show up in three 
representative cases: the top panel corresponds to $\alpha=1, \beta=1$, that is, to
the sum of Shannon entropies \cite{nuestro}; the middle panel corresponds to the `unspecial'
case $\alpha=2, \beta=2/3$; the bottom panel corresponds to $\alpha=1/2$, $\beta=\infty$.
Notice that in this latter case $R_\gamma^{(\rm inf)}=-\ln[ \max_p |\phi(p)|^2]$. Although the
three cases reported here yield similar results, it cannot be discarded that the freedom
in choosing, say, $\alpha$ may be used with advantage in some specific cases. 
The vertical, green solid line indicates again that collapse time-scales can be 
estimated within the entropic approach as the first maximum of the sum of entropies.

\begin{figure}
\includegraphics[width=8.5cm]{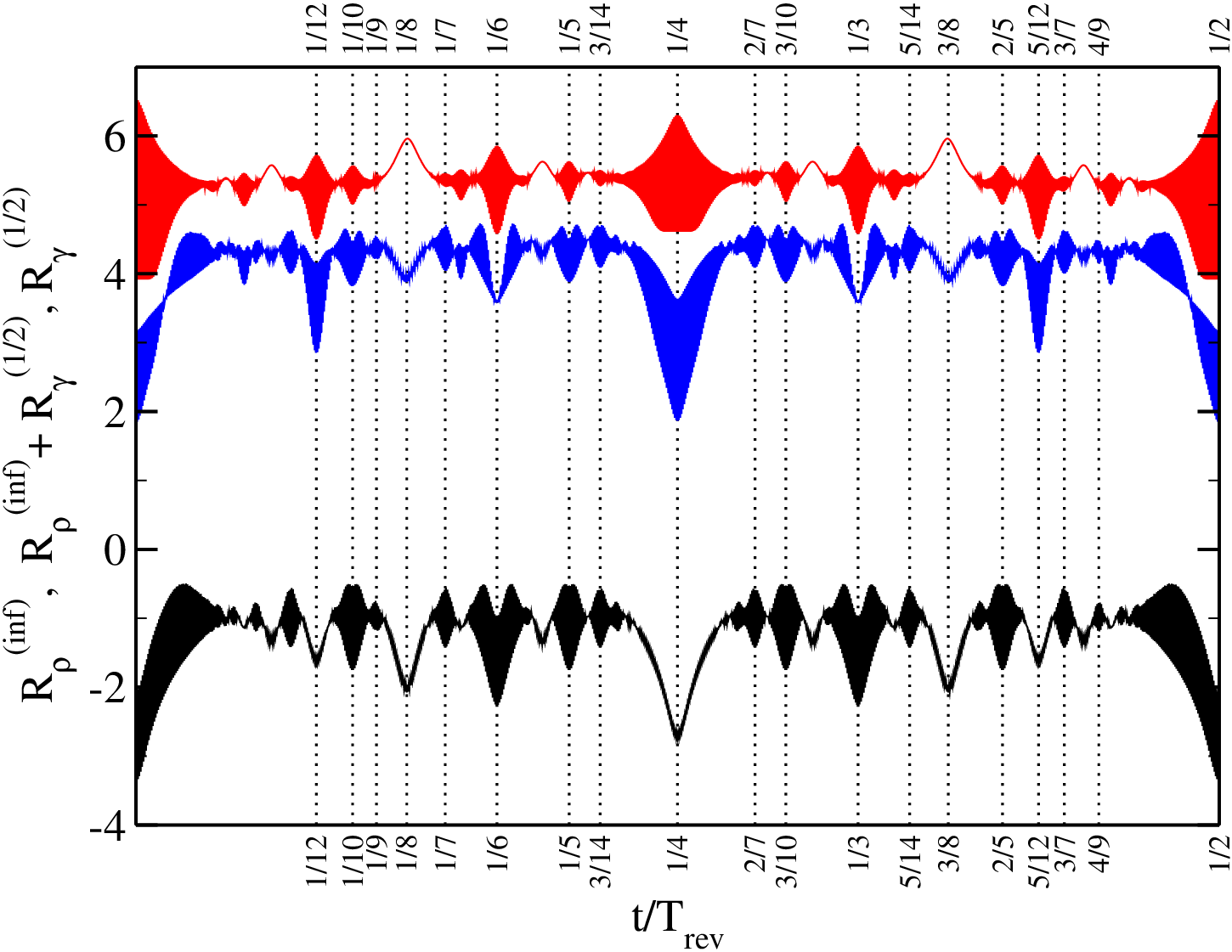}
\caption{(color online). Time dependence of (top red curve) $R_{\gamma}^{(1/2)}(t)$,
(middle blue curve) $R_{\rho}^{({\rm inf})}(t)+R_{\gamma}^{(1/2)}(t)$, and (bottom black curve) 
$R_{\rho}^{({\rm inf})}(t)$ for an initial Gaussian wave packet in an infinite square-well.
The main fractional revivals are indicated by vertical dotted lines.
Parameters as in Fig. \ref{fig1}} 
\label{fig3}
\end{figure}

In figure \ref{fig3} we illustrate the importance of taking the sum of the entropies
as an indicator of fractional revivals, and not either of them separately 
The top curve corresponds to $R_\gamma^{(1/2)}$, the bottom curve to 
$R_\rho^{(\rm inf)}=-\ln [\max_x |\psi(x)|^2]$ (which owes its wavy-looking
appearance to the presence of the $\max$ operator), and the middle one to their sum.
It is only the sum of the entropies that embraces 
both the configurational and dynamical 
aspects of the wave packet evolution through the uncertainty relation.

\subsection{Quantum bouncer}

Next, consider a particle bouncing on a hard surface under the
influence of gravity, that is, a particle in a potential
$V(z)=mgz$, if  $z>0$ and  $V(z)=+\infty$ otherwise.
The eigenfunctions and eigenvalues are given by \cite{gea}
\begin{equation}
\tilde{E}_n=z_n; \quad u_n(\tilde{z})= {\cal N}_n
{\rm Ai}({\tilde z}-z_n); \quad n=1,2,3,\ldots
\end{equation} 
where
$l_g=\left(\hbar/2gm^2\right)^{1/3}$ is a characteristic gravitational length,
$\tilde{z}=z/l_g$, $\tilde{E}=E/mgl_g$,  Ai$(z)$ is the Airy function,
$-z_n$ denotes its zeros, and ${\cal N}_n=|{\rm Ai'}(-z_n)|^{-1}$ \cite{vallee} 
is the $u_n({\tilde z})$ normalization factor. In the remainder of this paper 
the tildes on the energy and position variables will be omitted.
$z_n$ and $N_n$ were determined numerically by using 
scientific subroutine libraries for the Airy function,
although accurate analytic approximations for them can be found in \cite{gea}.
We have verified that the numeric approach yields slightly better results.
The corresponding coefficients of the wave function
can be obtained analytically as \cite{vallee}
\begin{equation}
a_n={\cal N}_n \left(2^{3/2}\pi \sigma\right)^{1/4} 
{\rm Ai}\left(z_0-z_n+\frac{\sigma^4}{6} \right) 
e^{\frac{\sigma^2}{2}\left(z_0-z_n+\frac{\sigma^4}{6}\right)}.
\end{equation}
Consider now an initial Gaussian wave packet 
localized at a height $z_0$ above the floor, with a width $\sigma$ and an initial 
momentum $p_0=0$. The classical period can be calculated to obtain $T_{\rm cl}=2\sqrt{z_0}$. 
As for the revival time, the state $n_0$ around which the initial wave packet is peaked 
can be identified via $z_0=E_{n_0}$ with the help of $z_n\backsimeq\left[3\pi\left(4n-1\right)/8\right]^{2/3}$ \cite{gea}.
One finds by direct substitution $T_{\rm rev}=2h/|E''_{n_0}|=8 z_0^2/\pi$, but it can bee shown that at a time half of this value
the wave packet reforms half of period out of phase with the classical motion (see \cite{gea}).
Therefore, we henceforth take $T_{\rm rev}=4 z_0^2/\pi$.
The temporal evolution of the wave packet in momentum-space was obtained numerically
by the fast Fourier transform method.

\begin{figure}
\includegraphics[width=8.5cm]{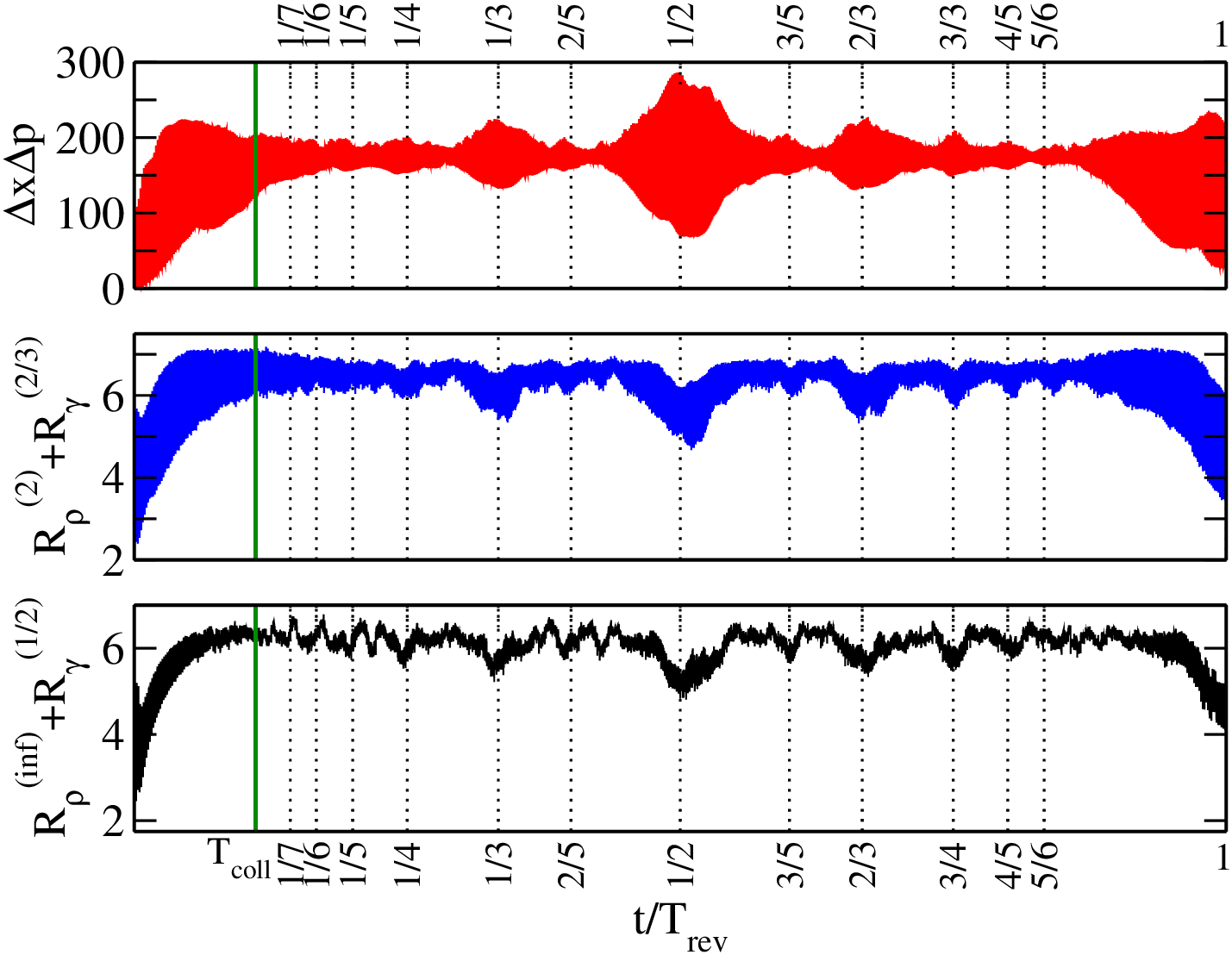}
\caption{(color online). 
Time dependence of (top panel, red curve) $\Delta x(t) \Delta p(t)$, 
(middle panel, blue curve) $R_{\rho}^{(2)}(t)+R_{\gamma}^{(2/3)}(t)$, 
and (bottom panel, black curve) $R_{\rho}^{({\rm inf})}(t)+R_{\gamma}^{(1/2)}(t)$ 
for an initial Gaussian wave packet with $z_0=100$, $p_0=0$, and
$\sigma=1$ in a quantum bouncer.
The main fractional revivals are indicated by vertical dotted lines,
and the vertical, green solid line stands for the collapse time.
}
\label{fig4}
\end{figure}

We have computed the temporal evolution of the uncertainty product 
$\Delta x \Delta p$ and the sums of R\'enyi entropies $R_\rho^{(2)}+R_\gamma^{(2/3)}$ 
and $R_\rho^{(\rm inf)}+R_\gamma^{(1/2)}$ for the initial conditions $z_0=100$,
$\sigma=1$ and $p_0=0$. Figure \ref{fig4} displays these quantities and the location
of the main fractional revivals. The top panel shows that $\Delta x(t)\Delta p(t)$
decreases and reaches a minimum at some fractional revivals. Analogous
behavior is seen in the middle and bottom panels for the sum of R\'enyi entropies.
The description in these  latter cases is somewhat more complete. 
Also shown in the same figure is the collapse time-scale 
$T_{\rm coll}={T_{\rm cl}^3}/(4\sqrt{2}\sigma)\simeq 1414.21$ located close to the maximum entropy
area \cite{gea}.

\section{Summary}
In this paper we have generalized and expanded on the entropic approach 
put forward in \cite{nuestro} by using R\'enyi uncertainty relations to 
study revival behavior in three model systems, namely, 
the simple harmonic oscillator, the infinite square well, and the quantum 
bouncer. We have found that they provide a useful framework for visualizing 
fractional revivals, alternative to the usual autocorrelation 
function and expectation value analyses.

Additionally, we have also successfully used the standard variance-based 
uncertainty product to search for the fractional revivals, but the information entropy turns out 
to be a more satisfactory measure of dynamical properties of wave packets than the moments of the probability 
distribution. We have also shown that collapse time-scales can also be computed within the 
entropic approach. In summary, we have shown that entropic uncertainty relations
are a good tool to investigate properties of the temporal evolution
of wave packets.

\section{acknowledgments}
This work was supported in part by the
Spanish projects FIS2005-00973 (Ministerio de Ciencia y
Tecnolog\'ia), FQM-2725 and FQM-165 (Junta de Andaluc\'ia).


\begin{thebibliography}{99}

\bibitem{1} 
J.H. Eberly, N.B. Narozhny, and J.J. S\'anchez-Mondrag\'on,
Phys. Rev. Lett. {\bf 44}, 1323 (1980).

\bibitem{1b} 
I.Sh . Averbukh and J.F. Perelman, 
Phys. Lett. A {\bf 139}, 449 (1989); 
Acta Phys. Pol. {\bf A78}, 33 (1990).

\bibitem{1c}
D.L. Aronstein and C.R. Stroud Jr.,
Phys. Rev. A {\bf 55}, 4526 (1997).

\bibitem{rob} 
R.W. Robinett, 
Phys. Rep. {\bf 392}, 1 (2004).

\bibitem{yeazell}
J.A. Yeazell, M. Mallalieu, and C.R. Stroud, Jr.,
Phys. Rev. Lett. {\bf 64}, 2007 (1990).

\bibitem{exp} 
G. Rempe, H. Walther, and N. Klein, 
Phys. Rev. Lett. {\bf 58}, 353 (1987); 
T. Baumert {\em et al.},
Chem. Phys. Lett. {\bf 191}, 639 (1992);
M.J.J. Vrakking,  D.M. Villeneuve, and A. Stolow,
Phys. Rev. A {\bf 54}, R37-R40 (1996);
A. Rudenko {\em et al.}, 
Chem. Phys. {\bf 329}, 193 (2006).

\bibitem{isotope}
I.Sh. Averbukh {\em et al.}, 
Phys. Rev. Lett. {\bf 77}, 3518 (1996).


\bibitem{factorization}
M. Mehring, K. M\"uller, I.Sh. Averbukh, W. Merkel, and W.P. Schleich,
Phy. Rev. Lett. {\bf 98}, 120502 (2007);
M. Gilowski, T. Wendrich, T. M\"uller, Ch. Jentsch, W. Ertmer, E.M. Rasel, and W.P. Schleich,
Phys. Rev. Lett. {\bf 100}, 030201 (2008);
D. Bigourd, B. Chatel, W.P. Schleich, and B. Girard,
Phys. Rev. Lett. {\bf 100}, 030202 (2008).

\bibitem{sun} 
C. Sudheesh, S. Lakshmibala, and V. Balakrishnan, 
Phys. Lett. A {\bf 329}, 14 (2004).

\bibitem{don}
M.A. Donchesk and R.W. Robinett,  
Am. J. Phys. {\bf 69}, 1084 (2001).

\bibitem{laserphysics}
D.L. Aronstein and C.R. Stroud, Jr.,
Laser Phys. {\bf 15}, 1496 (2005).

\bibitem{nuestro} 
E. Romera and F. de los Santos, 
Phys. Rev. Lett. {\bf 99}, 263601 (2007).


\bibitem{bouncer1}
K. Bongs {\it et al.},
Phys. Rev. Lett. {\bf 83}, 3577 (1999).

\bibitem{bouncer2} 
B.M. Garraway and K.A. Suominen, Contemp. Phys. {\bf 43}, 97  (2002); 
Ch. Warmuth {\it et al.}, J. Chem. Phys. {\bf 114}, 9901 (2001).

\bibitem{cite14} 
O. G\"uhne and M. Lewenstein, 
Phys. Rev. A {\bf 70}, 022316 (2004); 
G. Adesso, A. serafini, and F. Illuminati, 
Phys. Rev. A {\bf 70}, 022318 (2004); 
F.A. Bovino {\em et al.},
Phys. Rev. Lett. {\bf 95} 240407 (2005); 
B.M. Terhal, 
Theor. Comput. Sci. {\bf 287}, 313 (2002).

\bibitem{7}
P. Lay, S. Nagy, and J. Pipek, 
Phys. Rev. A {\bf 72}, 022302 (2005).

\bibitem{cite1416} 
D.G. Arb\'o {\em et al.}, 
Phys. Rev. A {\bf 67}, 063401 (2003);
S. Gnutzmann and K. Zyczkowski,
J. Phys.  A: Math. Gen. {\bf 34}, 10123 (2001);
F. Verstraete and J.I. Cirac, 
Phys. Rev. B {\bf 73}, 094423 (2006).

\bibitem{cite89} 
C. Beck and D. Graudenz, 
Phys. Rev. A {\bf 46}, 6265 (1992);
S. Kohler and P. H\"anggi, 
in Quantum Information Processing, edited by G. Leuchs and T. Beth  (Wiley-VCH, Berlin, 2002).


\bibitem{biarenyi} 
I. Bianilicki-Birula,
Phys. Rev. A {\bf 74}, 052101 (2006).

\bibitem{BBM} 
I. Bialynicki-Birula and J. Mycielski,
Comm. Math. Phys. {\bf 44}, 129 (1975);
W. Beckner, Ann. Math. {\bf 102}, 159 (1975).

\bibitem{kennard} 
E.H. Kennard, 
Z. Phys. {\bf 44}, 326 (1927).



\bibitem{bluhm} 
R. Bluhm, V. A. Kostelecky and J.A. Porter, 
Am. J. Phys. {\bf 64}, 944 (1996).

\bibitem{styer} 
D.F. Styer,
Am. J. Phys. {\bf 58}, 742 (1990).


\bibitem{4} 
D.L. Aronstein and C.R. Stroud, Jr.,
Phys. Rev. A {\bf 55}, 4526 (1997).

\bibitem{rob2}
R.W. Robinett,
Am. J. Phys. {\bf 68}, 410 (2000).

\bibitem{uffink}
H. Maasen and J.B.M. Uffink, 
Phys. Rev. Lett. {\bf 60}, 1103 (1988);
J.B.M. Uffink, 
Ph. D. Thesis, University of Utrecht, 1990 (unpublished)
(http://www.phys.uu.nl/igg/jos/publications/proefschrift.pdf).

\bibitem{gea} 
J. Gea-Banacloche, 
Am. J. Phys. {\bf 67}, 776 (1999).

\bibitem{vallee}
O. Vall\'ee,
Am. J. Phys. {\bf 68}, 672 (2000).

\end{thebibliography}
\end{document}